\documentclass[a4paper,twoside]{report}
\usepackage[english]{babel}
\usepackage{footmisc}
\usepackage{epsfig}
\usepackage{float}
\usepackage{amssymb}
\usepackage{amsmath}
\usepackage{subfigure}
\usepackage{graphicx}
\begin{document}
\title{Test and Developments of Crystals for a High-Resolution
Electromagnetic Calorimeter for PANDA} \author{Master Thesis by
\\Sophie Ohlsson \\ Supervisor: Ulrich Wiedner \\\\\\\\\\
Department of Radiation Sciences, Uppsala University}
\date{June 2004}
\maketitle

\cleardoublepage
\chapter*{Abstract}
The particle beam facility called GSI, located in Darmstadt (Germany),
will within a few years be upgraded with a new complex of
accelerators. At the new storage ring HESR, the PANDA experiment aims
to study antiproton induced reactions. The electromagnetic calorimeter
will constitute a major part of the experimental set-up. The detector
will be used to detect photons and is planned to consist of thousands
of scintillating crystals.

In this thesis, the energy resolution of crystals and its dependence
on the incoming energy has been investigated. The crystals used were
first prototype crystals of a new generation, characterized by a
higher light yield. Two crystal types with different shape and light
yield, manufactured by suppliers in China and Russia, were used.

An experiments were performed on two arrays of $3\times3$
crystals. Photon beams of eight different energies, ranging from
64~MeV to 715~MeV, were directed into the center crystals. Due to the
development of electromagnetic showers the energy was deposited over
all nine crystals. Photo Multiplier Tubes were used to read out the
response of the scintillators.

The energy peaks obtained in the experiment were not entirely
symmetric. The general line shape of the experimental energy response
was a Gaussian with an asymmetric tail towards lower energies due to
energy losses. From this line shape one can get a measure of the width,
sigma, of the peak. In this thesis, the sigma value was obtained using
three different methods:

\begin{enumerate}
\item FWHM (Full Width Half Maximum)
\item Two Gaussian method
\item Right side fit
\end{enumerate}

All three methods result in different energy resolutions and the goal
was to determine which method describes the course of events most
accurately.  

The FWHM method is used whenever one is dealing with non-Gaussian
distributions. The second method is a method in which the
asymmetric energy peak is represented by two different Gaussian
distributions. This method reproduces more accurate the line shape of
the energy peaks due to the energy losses caused by the wrapping
material and the statistical fluctuations of the amount of energy
escaping the limited detector volume. The right side fit is a method
where only the right side of the peak is fitted with a Gaussian
function and the tail on the left side of it is completely ignored.

The real energy resolution does not depend on what method is used in
describing it. It is however important to find a theory that well
describes the energy resolution of the crystals used in the
experiments. 

Calculations showed that using the two Gaussian method delivered the
lowest sigma value and also the lowest energy resolution. The FWHM
method gave the largest sigma value and also the largest energy
resolution. The right side fit gave an energy resolution with values
between the two other methods, thus this method was chosen to describe
the crystals. Error bars were estimated from the other two
approaches. The obtained energy resolution for the Russian crystals,
$1.61\%/\sqrt{E}+0.73\%$, was much better than the resolution for the
Chinese crystals, $1.58\%/\sqrt{E}+4.25\%$. It actually came close to
what was described in the Conceptual Design Report (CDR). However, the
energy resolution mentioned in the CDR, $1.54\%/\sqrt{E}+0.3\%$, was
obtained for a significantly larger $5\times5$ crystal array
containing most of the shower. This response function was determined
at a moderate operating temperature of +8 degrees~C, while the
measurements for this thesis were performed at -25~degrees~C. The
present result may be limited by a significantly thicker layer of dead
material between the crystal elements. Therefore the use of the signal
peak method might be justified to estimate the overall light output of
the crystal array. Separating the expected contribution due to energy
losses may very well be reasonable, due to the limited transversal
dimensions of the crystal array.

Some simulations were as well performed. These showed that the showers
spread beyond the nine crystals used in the experiment. This was
especially noticeable when incoming photons with a high energy were
used.

\cleardoublepage
\tableofcontents
\cleardoublepage

\chapter{Introduction}
\label{intro}
PANDA is an international collaboration. The acronym stands for
antiProton ANihilation DArmstadt and the objective for this
collaboration is to increase the understanding of the strong
force. The facility, at which the research is conducted, is located
at GSI (Gesellschaft f\"{u}r Schwerionen) in Darmstadt, Germany.

Over 30~years ago, the GSI laboratory was founded; the goal was to
build a research facility for nuclear physics and related areas. The
construction of a new heavy-ion accelerator soon attracted scientists
from all over Europe, and since the upgrade 15 years ago, GSI has
evolved into an international research centre using beams of
heavy-ions up to en energy of 2 AGeV.

Today there are plans for a major new research facility at GSI. The
new facility will provide a range of particle beams from proton and
antiproton beams to ion beams up to Uranium. The extension will mainly
consist of a new 100/200~Tm double-ring synchrotron referred to as
SIS100/200 and storage rings for collecting and cooling the beam as
well as for phase space optimization and experiments.

Research with antiprotons, hadron spectroscopy and hadronic matter
will be conducted in the new facilities to gain a greater knowledge of
the strong force. When colliding protons and antiprotons into each
other, particles with gluonic degrees of freedom are produced. Studies
of these particles, such as precision measurements of the mass, width
and decay branches of charmonium aim to seek answers to questions like
``Where does the mass of a nucleon come from?''. 

Charmonium will be studied carefully at GSI since the coupling
constant of this meson is not too large and relativistic problems are
considered to be manageable. Another interesting property of
charmonium is that the mass of the charm quark is large enough to
motivate the use of perturbative QCD (Quantum Chromo Dynamics) and
still small enough to keep the non-perturbative corrections. The
reaction chain in which the charmonium hybrid state $\psi_{g}$ is
created, will eventually end with only photons and electrons as the
final products, as shown below.\cite{GSI}
\newpage

\begin{eqnarray*}
p+\overline{p}&\rightarrow&\psi_g +\pi^0\\
\psi_g &\rightarrow&\chi_{c1}+(\pi^0\pi^0)_s\\
\chi_{c1}&\rightarrow&\gamma+J/\psi\\
J/\psi&\rightarrow& e^+e^-\\
\pi^0&\rightarrow& \gamma\gamma\\
\end{eqnarray*}

The reaction chain described above ends with seven gammas, one
electron and one positron. The gammas can only be detected using an
Electromagnetic CALorimeter (ECAL) of nearly 4$\pi$ solid angle
coverage. As charged particle detectors must be positioned closer to
the target than the electromagnetic calorimeter, the radius for this
detector grows considerably and thousands of scintillator crystals
will be needed. The electrons and positrons can be detected using
either an ECAL or a charge sensitive detector. It is required that the
ECAL has a high resolution and a sufficiently fast response to detect
all particles created in the decay chain.

My role in the PANDA-project has been to perform measurements on
Chinese and Russian crystals in Germany to investigate their energy
resolution and to determine if the crystals match the desired energy
resolution of $\frac{1.54\%}{\sqrt{E(GeV)}}+0.3\%$. 

This report treats both the physics behind the experiment and the
connection to the research pursued at GSI in Darmstadt. Firstly, the
reader will be guided through the theory and background to the
experiments. The reader will be introduced to the PANDA experiment as
well as the reason to, and function of, electromagnetic
calorimeters. The preparations in Giessen and the measurements in
Mainz will be presented in chapters \ref{giessen} and \ref{mainz}, as
well as the experimental set-up, methods used and the obtained
results. Finally, a summary and outlook will be presented.

\cleardoublepage
\chapter{Theoretical Background}
\label{theory}
\section{Fundamental Building Blocks}
The matter in nature is made up of two fundamental groups of building
blocks. These non-excited elementary particles all have spin 1/2 and
are called fermions. The fermions are either \textit{leptons} or
\textit{quarks}. In total, six leptons and six quarks have been
found. According to their properties, these have been classified in
three generations, or flavours. The leptons are called $e^-$, $\mu^-$,
$\tau^-$, $\nu_e$, $\nu_\mu$ and $\nu_\tau$. The first three leptons
carry the charge -e while the last three, called the neutrinos, have
no charge. Each lepton has the same flavour, or belongs to the same
generation, as its corresponding neutrino. In addition to these
leptons there exist six antiparticles. They are called $e^+$, $\mu^+$,
$\tau^+$, $\overline{\nu_e}$, $\overline{\nu_\mu}$ and
$\overline{\nu_\tau}$. The charged leptons interact via both the
electromagnetic force and the weak force, while the neutral leptons
interact weakly.

The six quarks occur in flavours consisting of two particles, just
like the leptons. The quarks are called up-, down-, strange-, charm-,
top- and bottom quarks. Each generation consists of a quark with
charge $\frac{-1}{3}e$ and a quark with charge
$\frac{2}{3}e$.\cite{MartinShaw} 

A single quark has never been found, this phenomenon is called
confinement. Instead, the quarks stick to each other in formations
called \textit{hadrons}. Hadrons consisting of two quarks are called
\textit{mesons}, hadrons consisting of three quarks are called
\textit{baryons}. 

To ensure that the quarks obey the Pauli principle, one must have a
way to discern them quark from another. The quarks have therefore been
allotted colour charges. They carry either red, blue or green colour,
while the antiquarks carry anti-red, anti-blue or anti-green.

The interaction binding the quarks into hadrons is called the strong
interaction and the mediator of this force is the \textit{gluon}. The
gluons couple to the colour charge carried by the gluons in the same
way as the electromagnetic force couples to the charge of particles. The
gluons carry colour and anticolour simultaneously.\cite{Povh}
\newpage

\section{Particle Interactions}
Particles interact via something called \textit{mediators}. A mediator
is a particle that is exchanged in the interaction process. To depict
the interactions in a pedagogical way, \textit{Feynman diagrams} are
often used. These are diagrams showing both time and space and each
symbol in this diagram corresponds to a certain matrix element. There
are certain rules connected to Feynman diagrams. The time axis usually
runs upwards, while the space axis runs from left to right. Different
kinds of lines characterize different particles in the interaction
process. Straight lines correspond to wave functions of fermions,
while antiparticles are depicted having arrows pointing backwards in
time. Photons have wavy lines, heavy vector boson have dashed lines
and gluons have corkscrew-lines.

The particles interact at the point called the \textit{vertex}. If
more than three particles meet, the points are called
\textit{vetices}. At this point energy, momentum and electric charge
must be conserved. Here, particles not being present in the initial
state often show up. These are called \textit{virtual particles} and
do not have to satisfy the energy conservation law since they live for
a very short period of time. 

For each vertex there is a transition amplitude that contains a factor
$\sqrt{\alpha}$. $\alpha$ is a dimensionless strength parameter which
is independent of the particle types involved in the process, though
it depends on what interaction is taking place. The $\alpha$ value for
the strong interaction is 1, while the values for the weak and
electromagnetic forces are $1^{-6}$ and 1/137
respectively.\cite{alpha} If there are three vertices in an
interaction process, the probability for this interaction to take
place is $\alpha^3$.\cite{Povh}

\section{Quantum Electrodynamics and Quantum \\ Chromodynamics}
Quantum ElectroDynamics, or QED, is the quantized, relativistic theory
of electrons and positrons in interaction with the electromagnetic
field. The electromagnetic force acts on large distances and the
mediator responsible for this interaction is the spin-1 boson called
the photon. 

Quantum ChromoDynamics, QCD, is the theory of the quark (and gluon)
interaction and it focuses on the strong force. This force, acting
between quarks, is much stronger than the electromagnetic force and it
also acts on much smaller distances. The particles responsible for the
interaction are in this case the eight massless spin-1 bosons called
gluons. These interact with the quarks in the nuclei.\cite{williams}

\section{Standard Model}
The four elementary interactions known are the strong force, the weak
force, the electromagnetic force and gravitation. The electromagnetic
force and the weak force have been combined into what is called the
electroweak force and one is working on a theory on how to combine the
remaining two. The \textit{standard model} is a name for the existing
model of elementary particles and their interactions. It comprises
both the theory of the electroweak interaction and quantum
chromodynamics. 

The force mediators are the gluons, the photons, the vector bosons
$W^\pm$ and $Z^0$ and possibly the graviton which has not yet been
discovered. Since the gluons couple to colour and carry colour as well
as anticolour, the gluons couple to themselves.\cite{williams}

\section{Gluons and Quarks}
Gluons attract other gluons because of the colour (or anticolour) they
carry. Instead of spreading out in space like the electric field lines
between charges, the gluons are ``trapped'' in something called a flux
tube that links the quarks. The harder one tries to separate the
quarks, the stronger the field in between them gets. It would simply
take an infinite amount of energy to completely separate the
quarks. The only known way to break the gluon flux tube is to take the
energy stored in the field and create two new coloured
quarks.\cite{article}

Coloured objects are never seen in nature, this applies to both quarks
and gluons and the phenomenon is called confinement. Combinations
e.g. of many gluons should on the other hand be possible to
observe. These could consist of many gluons combining to colour neutral
particles e.g. a blue/anti-red and a red/anti-blue gluon.  

Three jet events in electron-positron annihilations are a strong
indication for their existence. In such annihilation processes two new
quarks that spread in opposite directions are (thought to be)
created. As the quark and antiquark ``escape'' each other, new mesons
and baryons are created as jets. Therefore an odd number of jets must
originate from gluons.

Gluons seem to be able to emit other gluons as well. 

\section{Exotic Particles}
\textit{Exotic particles} is a name for hadrons other than mesons and
baryons, including particles with excited gluonic fields. The
particles with exotic gluonic fields are either so-called
\textit{glueballs} or \textit{hybrids}. The glueballs consist only of
gluons, while the hybrids consist of excited gluons as well as
quarks.\cite{elin}

\subsection{Glueballs}
The glueball is thought to have a very small radius - in the range of
$10^{-17}$m, a hadron has a diameter of approximately
$10^{-15}$m.\cite{radius} Since no free coloured objects exist,
glueballs must be colour neutral objects. The glueball is predicted to
consist of at least two gluons carrying for instance blue/anti-red and
red/anti-blue colour. Another option is three gluons combining to a
so-called white object. The three gluons making up the glueball would
then be for instance red/anti-blue, green/anti-red and
blue/anti-green. There is nothing preventing more complex formations
to exist either. Glueballs may also consist of more than a two or
three gluons as long as the resulting particle is colour neutral.

The existence of glueballs has not yet been proven though several
models predict them. Calculations using QCD are not easy to perform
(but have been done) and thus massive simulations using the most
modern model, called lattice QCD, have to be conducted. Using lattice
QCD simulations, time-space is regarded as a grid, called a
lattice. In these simulations quarks and antiquarks are connected by
lines. The results from these simulations show that gluons actually do
turn up.\cite{article}

Glueballs with exotic quantum numbers (quantum numbers that are
forbidden for ordinary mesons and fermion-antifermion systems) are
called \textit{oddballs}. Since they cannot mix with mesons they
should be easier to identify than glueballs.

\subsection{Hybrids}
The hybrids are made up of both quarks and excited gluons. An example
of this mixing could possible be a quark carrying red colour, an
anti-blue antiquark and a blue/anti-red gluon. This results in a
``white'' particle. 

When using lattice QCD to find indications of hybrids, one has found
an interesting property of the meson. This property occurs if the
lines linking the quarks and antiquarks in the simulation program are
not stretched out enough. What happens is that a slack in the line
between the particles arise and the configuration may start to spin
like a jumping rope. In these situations one can think of an extra
gluon being attached to the meson. This increases the energy and the
hybrid will be more energetic than the original meson. 

The detection of the charmonium hybrids in the decay channel discussed
in chapter \ref{intro} is one of the prime objectives of PANDA. This
hybrid consists of a charm quark, an anticharm quark and gluonic
excitations.

\cleardoublepage
\chapter{Future Experiments at GSI}
\label{panda}
Three very interesting things to study in the research area of hadron
physics are: confinement, the creation of mass and the search for new
forms of matter. The fact is that only about 2$\%$ of the nucleon
mass is made up of quark masses. The rest must be connected to the
kinetic energy and the interaction energy of the quarks making up the
nucleus. 

To fully understand the three phenomena mentioned above, physicists
need a greater understanding of the strong force. They plan to collide
protons with antiprotons and thereby create new, short-lived
particles. One possibility is to create charmonium, a meson consisting
of a charm quark and an anticharm quark. Scientists thereby hope to
gain some understanding to the strong force and find evidence of the
existence of glueballs and hybrids.\cite{panda}

\section{Future Upgrade of the GSI Facility}
Today, the facility at GSI consists of a LINAC (LINear ACcelerator), a
heavy-ion synchrotron SIS18 and an Experimental Storage/cooling Ring
(ESR). The new facility will have two separate synchrotron accelerator
rings, both with a circumference of about 1100m. The new facility
will as well be complemented by three additional storage rings; the CR
(Collector Ring), the NESR (New Experimental Storage Ring) and the
HESR (High-energy Storage Ring). 
\newpage

\begin{figure}[H]
\begin{center}
\includegraphics[width=0.7\linewidth,angle=0]{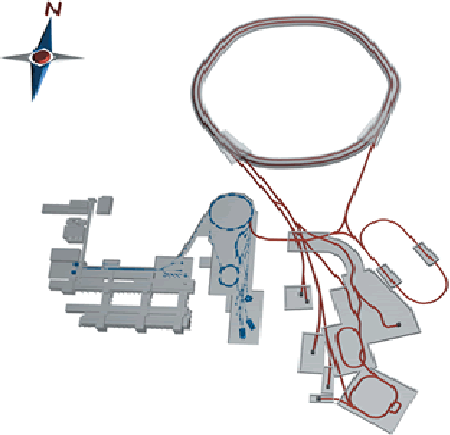}
\end{center}
\caption{The figure shows the present GSI facility to the left (marked
with blue colour). The planned upgrade of the facility is seen to the
right (marked with red colour).\cite{gsipage}}
\label{fig:newfac}
\end{figure}

The new detectors needed at GSI have to be of top quality to identify
the correct masses, nuclear charges, energies, momentum, angles
etc. The detector for the proton-antiproton collisions must be
designed to accept as many as $2\cdot10^7$ annihilations per second
and simultaneously allow for triggering on rare events with very low
cross-section. 

For complete information on e.g. spectroscopy measurements and the
reconstruction of invariant masses from neutral and charged decay
products, a nearly full coverage of the solid angle is required. For
particle identification it is as well very important to have excellent
energy and angular resolution, both in the case of charged particles
and for photons.\cite{GSI}

\section{Charmonium in the PANDA collaboration}
The reason for studying charmonium, instead of any other meson, in the
PANDA collaboration is that the charmonium states are very narrow and
separated. The cross-sections for the prediction of these states are
high, the relativistic effects are small and perturbative methods can
be applied. By analogy one can say that the charmonium is ``the
positronium of QCD'' \cite{poster}. By that one means that both
positronium (the bound state of an electron and a positron) and
charmonium are systems consisting of a particle and its antiparticle,
the difference is the force acting between them.\cite{Povh}

\begin{figure}[H]
\begin{center}
\includegraphics[width=0.7\linewidth,angle=270]{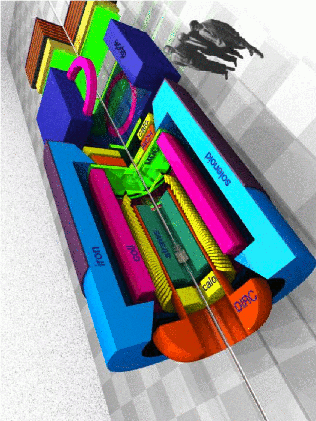}
\end{center}
\caption{The figure shows the PANDA detector which will be built at
GSI. The electromagnetic calorimeter is marked with yellow colour in
the figure.\cite{GSI}}
\label{fig:PANDA}
\end{figure}

Charmonium states often end up in an electron-positron final
state. These states are sometimes accompanied by photons and to detect
these photons, a new and powerful electromagnetic calorimeter is
needed. An illustration of the proposed PANDA detector (including an
electromagnetic calorimeter) labeled ``calo'' in yellow in figure
\ref{fig:PANDA}.

\cleardoublepage
\chapter{Electromagnetic Calorimeter}
\label{ecal}
\section{Principles of an Electromagnetic Calorimeter}
ECALs are used to detect and measure the energy and impact point of
photons, electrons and positrons with energies above 100~MeV. The
detection principle is based on registering the total absorption of
the electromagnetic shower \cite{williams}, which comprises
secondary particles created in bremsstrahlung and pair production
processes inside the detector.

Calorimeters can be constructed as homogeneous devices or in a
sampling mode, when only part of the shower energy is detected in
active detectors and the major part just absorbed in high density
material. The optimum resolution can be obtained for homogeneous
calorimeters using liquid noble gases or high density and fast
scintillation crystals. The ECAL of PANDA is supposed to be built with
scintillating crystals covering all showers. The close arrangement of
crystals will detect the full energy of the cascade created by an
incoming photon. 

The crystals produce visible wavelength photons when incoming photons
excite the detector material. The produced light can then be converted
into electrical signals using photo sensors such as photo multiplier
tubes or avalanche photo diodes.

The ECAL detects both the energy and the momentum of the particle. The
resolution of the momentum measurements decreases linearly with the
momentum of the particle while the energy resolution increase as a
function of $\frac{1}{\sqrt{E}}$. 

To identify the detected particle, mass and charge are in most cases
sufficient. The momentum is obtained by deflecting the particle in a
magnetic field and reconstructing the track in a position sensitive
detector volume consisting of straw tubes or gas detectors such as a
TPC (Time Projection Chamber). The mass is usually obtained by
additional measurements.\cite{Povh}
\newpage

\section{Energy Losses}
In general, the interaction of the incoming photons with the detector
material depends strongly on the energy of the incoming photons. If
the incoming photons have a high energy, this interaction process will
generate secondary particles which themselves generate new
particles. 

A \textit{shower} or a \textit{cascade} will be formed in cases where
the incoming photon energy is high. Initially, a phenomenon called
pair production will occur and when the energy has decreased,
processes like the photoelectric effect and Compton scattering will
take over. The shower process spreads in all directions but
prominently in the longitudinal one \cite{MartinShaw}.

\subsection{Bremsstrahlung}
The shower is generated by high-energy electrons or photons which lose
energy due to \textit{bremsstrahlung} when the particle interacts with
the detector material and photons are emitted. 

Translated into English ``Bremsstrahlung'' becomes braking
radiation. \\Bremsstrahlung occurs as particles are accelerated and
slowed down by the electric field from the nuclei in
matter.\cite{MartinShaw}

The amount of energy lost in bremsstrahlung depends on the mass and
charge of the particle according to equation \ref{eq:brems} \cite{ph}.

\begin{equation}
\label{eq:brems}
\frac{dE}{dx}=\frac{E}{X_0}
\end{equation}

$dE/dx$ is the lost energy, E is the energy of the particle and $X_0$
is the distance traveled by the particle.  At medium energies only
electrons and positrons have such low mass that the cross-section
becomes significantly large.

\subsection{Shower Process}
\textit{Pair production} is the interaction process between incoming
particles and matter that requires the highest energy
\cite{MartinShaw}. In this process the incoming photon converts
into an electron-position pair. Bremsstrahlung then occurs and the
newly formed electron (or positron) will be deviated and a photon will
be emitted. 

This shower, or cascade process, continues until the critical energy
$E_c$ of the particles is reached. At this electron energy, the
cross-section of bremsstrahlung becomes similar to that of pure
ionization. Therefore, no further secondary photons are generated and
the shower is stopped. The critical energy depends on the proton
number (atomic number) Z of the detector material.\cite{MJRyans89} A
material with a high Z-value, such as lead tungstate, corresponds to a
low critical energy.

\begin{equation}
\label{eq:crit}
E_{c}\approx\frac{550~MeV}{Z}
\end{equation}

At energies below a few~MeV, losses are dominated by the
\textit{photoelectric effect} and \textit{Compton scattering}. In the
photoelectric effect, the photon is absorbed by the atom as a whole
and an electron carrying the energy is emitted. In Compton scattering
the photon scatters off from an atomic electron. The photoelectric
effect and Compton scattering are both processes in which the photons
lose energy by collisions with the atomic electrons. This in turn
causes ionization of the atoms in the material. 

Concerning the performed experiments in Mainz, pair production was the
dominant interaction process for the incoming high energy photons,
while the photoelectric effect and the Compton scattering were
dominant in the end of the shower processes.

\section{Radiation Length}
In connection to the showers, the radiation length should also be
mentioned. The radiation length is physically described as the
distance over which the energy of an incoming electron decreases by a
factor of $1/e$ \cite{leo}. For a photon, this distance can be said to
correspond to the distance over which there is an approximate
probability of 54\% for a $\gamma$-ray to perform pair production.

The radiation length $X_0$ has an asymptotic value for each material
at sufficiently high energies and is given by equation
\ref{eq:radlen}.\cite{MartinShaw}

\begin{equation}
\label{eq:radlen}
\frac{1}{X_{0}}=4(\frac{1}{mc\hslash})^2Z(Z+1)\alpha^3n_{a}\ln{\frac{183}{Z^\frac{1}{3}}}
\end{equation}

In \ref{eq:radlen}, m represents the mass of an electron, c is the
speed of light, $\hslash$ is Planck's constant divided by $2\pi$, Z is
the atomic number, $\alpha$ is the coupling constant and $n_a$ is the
density of atoms per $cm^3$ in the material. 

The radiation length for $PbWO_4$ crystals is approximately 0.9cm
\cite{proceed}. Materials with a high Z-value have a short radiation
length. This is a desired property since it decreases the calorimeter
depth and reduces the overall dimensions of the detector. This is
particularly important if all components have to be installed inside a
superconducting solenoid, such as in case of PANDA.

\section{Electromagnetic Calorimeter at GSI}
Di-electrons ($e^+e^-$ pairs) and photons will be detected with an
electromagnetic calorimeter made of $PbWO_4$ crystals. The PANDA
electromagnetic calorimeter will consist of four parts: the barrel,
the backward endcap, the forward endcap and the forward
spectrometer. The inner part of the barrel will have a diameter of
50cm and it will contain more than 9000 crystals. Both endcaps will
consist of approximately 4-5000 crystals. The final geometry and
granularity has not been fixed yet. Simulations however suggest a
minimum depth of 20cm and a typical cross-section of 2cm, similar to
the Moliere radius of the crystals. The Moliere radius is quantity
that describes the electromagnetic interaction properties. It is often
used when describing the transversed dimensions of electromagnetic
showers in a material \cite{moliere}.
\newpage

\begin{figure}[H]
\begin{center}
\includegraphics[width=0.4\linewidth,angle=270]{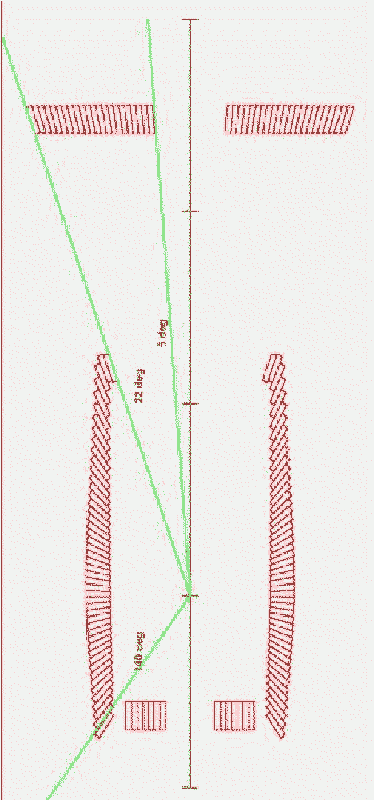}
\end{center}
\caption{In this figure the three central parts of the electromagnetic
detector are seen. The forward and backward endcaps are seen to the
right and left part of the figure respectively. The barrel is seen in
the middle of the figure.}
\label{fig:ecalparts}
\end{figure}

The forward spectrometer is located 7 meters downstream of the
target. It will have an approximate area of 3 $m^2$.\cite{fsemc}

\section{Crystal Properties}
The $PbWO_4$ crystals have some very attractive features: they are
fast scintillating crystals with a short decay time $<$ 10ns, a short
radiation length (0.9cm) and a Moliere radius of 2.2cm. The short
decay time allows experiments at high count rates and a fast
digitization even for developing low level trigger information
\cite{deadtime}. 

However, the luminescence yield correspond to only approximately 1\%
of NaI(Tl) crystals. Due to thermal quenching, the luminescence yield
can be significantly increased by operating the crystals well below
room temperature.
\newpage

\begin{figure}[H]
\centering
\includegraphics[width=0.6\linewidth,angle=270]{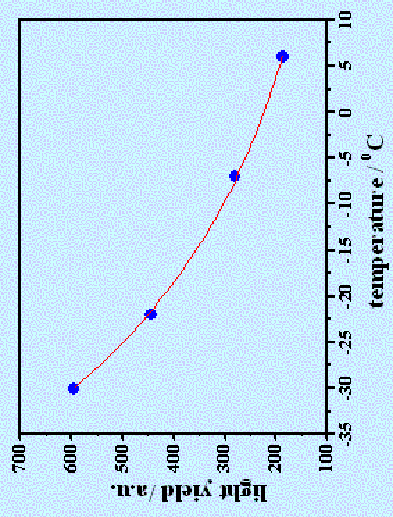}
\caption{The figure shows how the light yield of the lead tungstate
crystal varies with temperature. One can improve the light yield
multiple times by operating the crystals at low temperatures.}
\label{fig:lightyield}
\end{figure}

The energy resolution of the crystals, $\sigma/E$, is proportional to
$1/\sqrt{N}$ where N is the number of detected photons. Due to the
possibilities to increase the light yield of the crystals, a detailed
research program has been started to significantly improve it. 

An envisaged high granularity enhances the detector resolving power
with respect to position reconstruction and a minimized pile-up
probability.  The $PbWO_4$ crystals are in addition very compact since
their radiation length is short. This is a desired feature since a lot
of money and space can be saved by building smaller detectors. 

The crystals used in the measurements in Mainz were grown by the
company SIC at Shanghai, China and the Bogoroditsk Techno-Chemical
Plant, Russia. The Chinese crystals were tapered and had a length of
12cm, foreseen as prototypes for the proposed Photon Ball at COSY,
J\"{u}lich, Germany. As a rule of thumb, the crystals should have a
length of 15-20 radiation lengths and the dimensions of these crystals
may therefore indicate bad results. The light yield of the Chinese
crystals was approximately 10 photoelectrons/MeV. The second array
consisted of recently grown crystals from SIC as well as
Bogoroditsk. Both kinds had the same dimensions,
$2\times2\times15cm^3$ and a significantly improved light yield. The
Russian samples were produced with improved growing technology and
delivered a light yield nearly twice as much as the mass produced
crystals delivered for the CMS calorimeter at CERN.
\newpage

\begin{table}[htb]
\centering
\begin{tabular}{|l||c|c|}
\hline
Property &Chinese &Russian\\
\hline
\hline
Length [cm] &12 &15\\
\hline
Forward Dimension [$cm^2$] &$3\times3$ &$2\times2$\\
\hline
Backward Dimension [$cm^2$] &$2\times2$ &$2\times2$\\
\hline
Decay time [ns] &20 &20\\
\hline
Light Yield [phe/MeV] &10 &20\\
\hline
\end{tabular}
\caption{The table shows the approximate properties of the Chinese and
Russian crystals used in the experiment in Mainz. The correct
dimensions of the Chinese crystals can be seen in figure
\ref{fig:chrystals}.}
\label{tab:property}
\end{table}

\cleardoublepage
\chapter{Preparatory Measurements in Giessen}
\label{giessen}
\section{Crystal Preparation}
Normally, the crystals do not need any special preparation apart from
attaching the read-out devices at the back of each crystal. However,
in this case two different types of crystals were used. The older
Chinese crystals had already been wrapped individually in various
materials and in order to have as similar set-ups as possible, it was
decided that also the new Russian crystals were to be wrapped in the
same way. This individual wrapping procedure is something that
increases the dead material in the detector and worsens the energy
resolution and this should be avoided in the final construction of the
electromagnetic calorimeter. 

Firstly, the crystals were wrapped in eight layers of Teflon sheet
(total thickness approximately 0.1mm) as a diffuse
reflector. Secondly, they were covered by an Aluminum foil (0.15mm),
which in addition to its reflectivity provides an almost light tight
coverage of the crystal. 

The photo multipliers (Philips XP1911) were for electrical
insulation covered with a thin Kapton foil and a $\mu$-metal shield against
low magnetic fields. Additional wrapping with black tape minimizes
leakage of direct light from outside. 

Before attaching the PMT to the crystal, the side of the crystal that
would later be in contact with the PMT was cleaned using methanol. The
optical coupling was performed with Baysilone silicon oil of high
viscosity. The complete detector was assembled by a black plastic
shrinking tube, which provides sufficient mechanical stability as well
as makes the detector light tight.

\section{Experimental Set-Up}
\subsection{Crystal Set-Up}
\begin{figure}[H]
\centering
\includegraphics[width=0.6\linewidth]{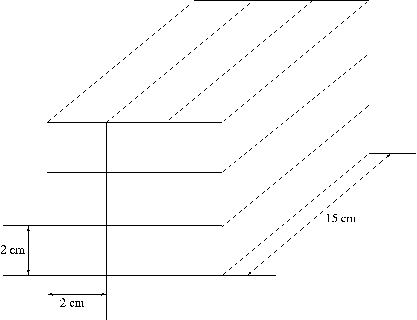}
\caption{A schematic figure of the $3\times3$ array with shown dimensions for
the Russians crystals.}
\label{fig:crystals}
\end{figure}

In total 18 crystals were arranged in two sets of 3x3 crystals, as
shown in figure~\ref{fig:crystals}.

\begin{figure}[H]
\centering
\includegraphics[width=0.9\linewidth]{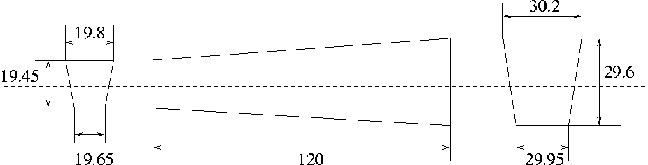}
\caption{Figure of the dimension of the tapered Chinese crystals used in the experiment. The dimensions are in mm.}
\label{fig:chrystals}
\end{figure}

One of the arrays consisted of older crystals grown by SIC,
China. They were tapered and had the dimensions shown in figure
\ref{fig:chrystals}.

In the second array, consisting of Russian as well as Chinese
crystals, the five samples of $PbWO_4$-II from Bogoroditsk were
arranged like a cross and the four SIC crystals were put in the
corners. 

A block of metal surrounded the two sets of crystals. Through this
block, cooling liquid circulated to enable measurements at low
temperatures (-24 deg~C) and to keep the crystals at a constant
temperature with a variation $<$ 0.1 degrees. The cooling of the crystals
is very important since the light yield is temperature dependent. In
fact, the light yield for $PbWO_4$ crystals improves by approximately
2-3\% for each lowered degree \cite{frida}. Five different
Platinum-resistive thermo sensors were integrated into the set-up to
control the cool-down phase, the reached final temperature and the
stability. 

The crystals were mounted on a table that could be moved by remote
control both horizontally and vertically to direct the beam into any
of the crystals elements. At the time of the measurements, the
crystals were surrounded by nitrogen gas to prevent formation of ice
on the electronic equipment.

\begin{figure}[H]
\centering
\includegraphics[width=0.6\linewidth,angle=270]{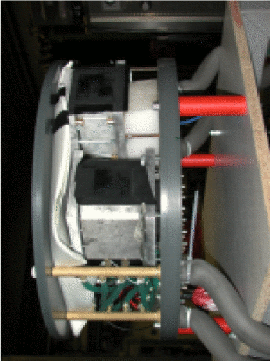}
\caption{The set-up used in Mainz showing the two arrays of crystals
contained in an aluminum housing to allow cooling. To the left the
block of shorter tapered crystals produced by SIC is seen, to the
right the array of new generation straight signals provided by both
suppliers.}
\label{fig:setup3}
\end{figure}

\section{The Muon Test}
In Giessen, the equipment and crystals were checked for failures to
ensure that everything would work during the beam time in
Mainz. Beside the use of low energy gamma sources, a simple way to
check the equipment is to try and detect cosmic muons. The muons are
minimum ionizing particles which means that they travel through mediums
with a minimum and constant loss of ionizing energy, approximately 13~MeV/cm for $PbWO_4$ \cite{link}. Due to the constant
energy deposition, these particles are advantageous to use even for
absolute energy calibration of an electromagnetic calorimeter at lower
or medium energies.\cite{link2}. 

Muons are very fast traveling particles, produced in air showers in
the atmosphere and have a large range into the Earth. They will
sometimes cross one, or maybe several of the uppermost detector
crystals (top layer) and then continue through one or several of the
bottommost (bottom layer) depending on the direction of the
trajectory. 

\begin{figure}[H]
\centering
\includegraphics[width=0.8\linewidth]{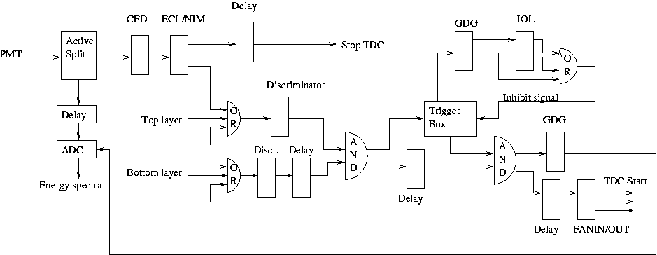}
\caption{Electronic scheme of the set-up. The two ORs have as input
the signals from the top layer crystals and the bottom layer
crystals, respectively. This set-up was used to detect muons crossing
crystals in both layers.}
\label{fig:hardware}
\end{figure}

By connecting logic signals which indicate a muon transition, from the
top and the bottom layer crystals in an logic AND-circuit, one can
select such events for the trigger. The measurement is performed by
creating such a trigger and recording events as list mode data. The
layout of the electronics scheme, based on commercially available NIM-
and CAMAC-electronics, as well as the trigger generation are shown in
figure \ref{fig:hardware}. A valid coincidence trigger enables the
start of the energy and time measurement relative to the trigger
reference, and offline true coincidences are selected among the
crystals. 

The used Data Acquisition System DAX was developed based on the TAPS
(Two Arms Photon Spectrometer) read-out system. In parallel it stores
the data of all detectors event-wise on disk and generates the relevant
spectra of the ADC as well as TDC information.

The PMTs (Philips XP1911) attached to the crystals, were supplied with
high voltage $<$ -1500V. The anode signals were connected to an Active
Split, which distributes the incoming signal into two or more
identical ones. In this case two outputs were used. 

One output without any applied shaping is passively delayed by coaxial
cables for 500ns before being fed into the charge sensitive CAMAC
Analog to Digital Converter, ADC (LeCroy 2249W).

The other signal from the Active Split continues to a Constant
Fraction Discriminator (CFD) to generate a timing reference. A CFD
gives an output if the input signal is above a certain level, the time
of this output is independent of the amplitude for a given rise time
of the input signal. To adapt the output to the NIM based electronics
the delivered ECL signal has to be converted in an ECL/NIM
converter. One of the two outputs is delayed in an active module for
390ns to serve as a stop signal in the Time to Digital Converter
(TDC), to make sure that it comes after the start signal, which is the
event trigger. The other signal acts as an input signal to a logic OR,
together with the two other signals from the top layer crystals. 

The following discriminator is used to adjust the logic output of the
OR to an appropriate width. The lower row of crystals in the array was
handled in a similar way to generate an OR information as well. A
short additional logic delay is used to make sure that always the
bottom row triggers and determines the time in case of a coincidence of
both rows in the AND module. 

The AND has two outputs. One of these outputs is actively delayed and
connected to a second AND. The second output is sent through a trigger
box, which selects acceptable events gated by an inhibit signal
generated in the crate controller. It considers the conversion time in
the ADC, the dead-time due to data transfer or the general busy
status. The accepted event is transfered to the crate controller via
the IOL-box (Input Output Latch-box), which is located in the CAMAC
crate. In order to correlate the accepted event trigger with the
original AND a second overlap is required in the circuit finally.

This second AND has two outputs. One of them enters a Gate and Delay
Generator (GDG), which provides the appropriate delay and width of a
NIM signal to gate the charge sensitive ADC correlated with the
corresponding analog signal. The second output is connected to an
active delay and then to a FANIN/FANOUT unit. The identical outputs
provide the start signal to the TDC. 

The test at the laboratory has been used to verify that all detectors
and electronic channels have been properly functioning. In addition,
it has been possible to adjust roughly the photo multiplier bias based
on the recorded dynamic range of the energy spectra. The most probable
energy deposition of the cosmic muons amounts to approximately
25-30~MeV.

\cleardoublepage
\chapter{Measurements at MAMI in Mainz}
\label{mainz}
\section{Mainz Microtron}
The energy resolution measurements were conducted at MAMI (MAinz
MIcrotron) in Mainz, Germany. It is an 855~MeV electron beam facility
that is being used to study nuclear and hadronic systems from an
electromagnetic point of view. The facility is being upgraded at the
moment and will deliver electrons up to an energy of 1.5~GeV by the end of
2005. The quality of the beam is outstanding with respect to the
emittance, reliability and stability.\cite{mami} 

In the experiments conducted at MAMI, the same crystal set-up as in
Giessen was used. The difference was a fast plastic scintillator
paddle, sized $15\times15cm^2$ and 10mm thick, that was mounted in
front of the crystal arrays. It was used to identify charged electrons
or positrons that may have been created upstream by pair production of
bremsstrahlung photons in the air. The signals from the paddle were
directly used in anti-coincidence mode with the trigger signal of the
crystal array to make sure that only events due to photons were
recorded. The plastic scintillator has due to its low Z-material a
very low probability for interaction with high energy photons
corresponding to $<$ 0.1\% $X_0$. The electronic read-out equipment
was very similar to the overall scheme for the test with cosmic
muons. Only the generation and selection of the event has been
modified.

Each tagger channel delivered a logic timing signal. The selected
eight channels corresponding to photon energies distributed over the
full energy range up to approximately 800~MeV were fed into an OR unit
after refreshing the signal in a discriminator unit. The event
condition requires a coincidence between the tagger-OR and the trigger
signal of the crystal module to be tested. The timing of the
coincidence was adjusted in such a way that the $PbWO_4$ crystal
determines the time and therefore guarantees the time relation for
energy and time measurement. The beam intensity was reduced to a value
to keep the count rate in the investigated $PbWO_4$ crystal well below
20~kHz to minimize pile-up as well as not to limit the resolution in
the AC-coupled charge sensitive ADC. 

The program of the measurement covered the relative calibration as
well as the final response measurement of both arrays. The calibration
was performed by shooting the direct photon beam, which has a diameter
of $<$ 5mm at the detector position, into each individual crystal to
roughly adjust the dynamic range in the ADC. The adjustments were made
with an appropriate photo multiplier voltage to store calibration data
for the energy deposition of the selected eight photon energies in the
exposed crystal. The remote controlled table allows for positioning of
each crystal, with high precision, into the beam axis.  After the
calibration a long data run was taken with the beam hitting the
central detector of each $3\times3$ array to measure the shower
distribution. All measurements have been performed at an operating
temperature of -24~degrees~C.

\section{Experimental Set-Up}
\subsection{Tagging the Photon Energy}
The photon beam used in the experiments is generated from an electron
beam hitting a radiator. When hitting the radiator, bremsstrahlung
occurs and the electrons emit photons. The electrons are deflected by
a strong magnetic field of a dipole magnet; electrons with higher
energy are bent less than electrons with a lower energy. Electrons of
the same momentum (or the same energy) will cross the so called focal
point of the magnet system irrespectively of the angle in which they
scatter. There are different focal points for different values of the
momentum and these points are located along a plane called the focal
plane, in which a ladder of nearly 300 small scintillation detectors
are located. By recording the responding tagger module in coincidence
with the response of the bremsstrahlung photon observed in the test
detector, one knows the exact photon energy. The typical uncertainty
of 1-2~MeV is given by the momentum coverage of the individual tagger
element in the focal plane. 

In this experiment eight different energies of the tagged photons in
the interval of 63 to 715~MeV were selected. The energies used were:
63.82~MeV, 105.60~MeV, 269.85~MeV, 313.14~MeV, 420.26~MeV, 520.78~MeV,
656.19~MeV and 714.53~MeV.

\subsection{Energy Deposits in the Crystals}
The nine crystals used are arranged in two arrays of $3\times3$
crystals. When directing the photon beam towards a crystal, the
photons initiate an electromagnetic shower that spreads over the
neighboring modules.

The energy deposits in the surrounding crystals depend on the photon
energy and can be reduced due to the dead material such as Teflon,
tape and plastics between the crystals.  The final goal therefore has
therefore been to measure the energy depositions in the nine crystals
simultaneously.

\section{Analysis of the Experimental Data}
In oder to deduce the line shape of the detector matrix, the energy
deposit in each crystal due to the shower has to be summed-up
event-wise. This means that for each event, one has to find the amount
of energy deposited in crystal number 1, crystal number 2 and so
on. To be able to do this, a relative calibration must first be
performed to ensure that a certain channel number in one crystal
corresponds to the same channel number in all other crystals. 

To be able to completely analyze the data and obtain a result for the
energy resolution, three different measurements had to be done: the
pedestal determination, the calibration procedure and the final
response measurement. The pedestal is the channel number in the
digitized energy spectra, which corresponds to a photon energy equal to
zero. This information is needed for the relative as well as absolute
calibration. The pedestal measurements, unlike the other measurements,
requires no beam.

\subsection{Pedestal Measurement}
The pedestal marks the zero point energy for each digitized spectrum
of the crystal response. Using a charge sensitive ADC, the pedestal
corresponds to the integral over the noise along the adjusted gate
width. The experimental gate width was adjusted to 600ns in spite of
the fast $PbWO_4$ response delivering at least 95\% of the
scintillation light within 100ns. However, since the data acquisition
system had to be located outside the experimental area, 30m of coaxial
cables had to be used to transfer the analog signals. In addition, the
adjustment of the coincidence with the tagger channels as well as all
the additional logic decision required a passive delay of 500ns,
achieved with coaxial cables as well. As a consequence, the long
cables caused a strong damping of the signal amplitude and a
significant increase of the effective signal width.

The pedestals are different for each detector channel. These pedestal
events have been generated by triggering the data acquisition with
uncorrelated pulser signals leading to a random integration over the
noise. To get the position of this pedestal, a histogram of the energy
data was created. Since the histogram shows a narrow, symmetric
pedestal peak the mean value can be taken as the position of the
pedestal. Alternatively, a Gaussian curve could have been fitted to
each peak to determine the position.

\subsection{The Response Measurement}
During the response measurement the photon beam was directed into the
central crystal of both arrays, recording the data from the eight
neighboring crystals as well simultaneously. The energy information,
the time information and the tagger information were recorded and
stored event-wise. The time information delivers the relative time of
response of the neighboring crystals to separate off-line random
coincidences. 

\begin{figure}[H]
\begin{center}
\subfigure[Chinese crystals]{
\includegraphics[width=0.6\linewidth,angle=270]{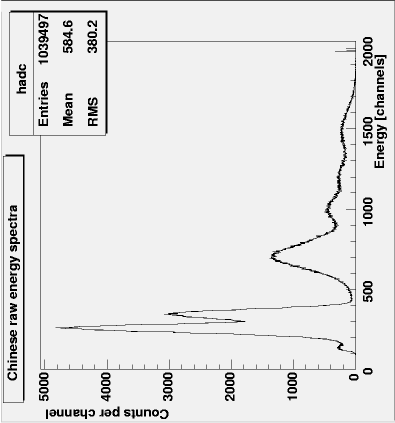}
\label{fig:rawch}}
\subfigure[Russian crystals]{
\includegraphics[width=0.6\linewidth,angle=270]{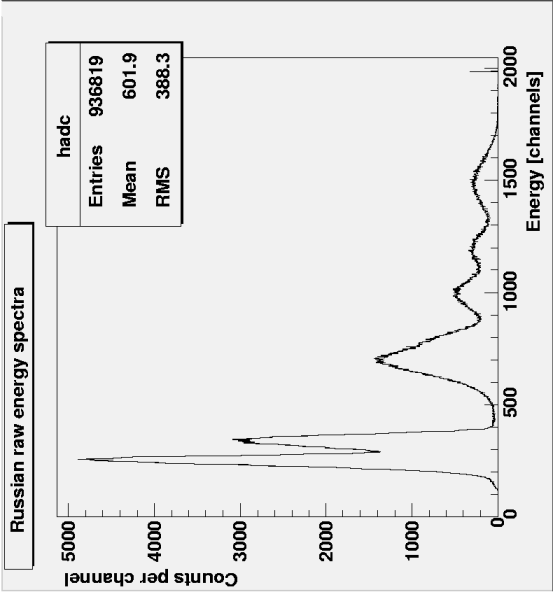}
\label{fig:rawru}}
\caption{Figures of the uncalibrated energy spectra from the Chinese
and Russian center crystal. The energy on the xaxis is shown in
channels.}
\end{center}
\end{figure}

The total energy spectrum of the center crystal shows eight energy
peaks since the beam is directed into that crystal; one peak for each
tagger used in the measurement. The eight other spectra show only low
energy deposit according to the fraction of the showers reaching these
crystals.  In \ref{fig:rawch} and \ref{fig:rawru} the eight tagger
peaks are shown for the center crystal. These spectra will in the
future be referred to as the raw energy spectra.

\begin{figure}[H]
\begin{center}
\subfigure[Chinese crystals]{
\includegraphics[width=0.6\linewidth,angle=270]{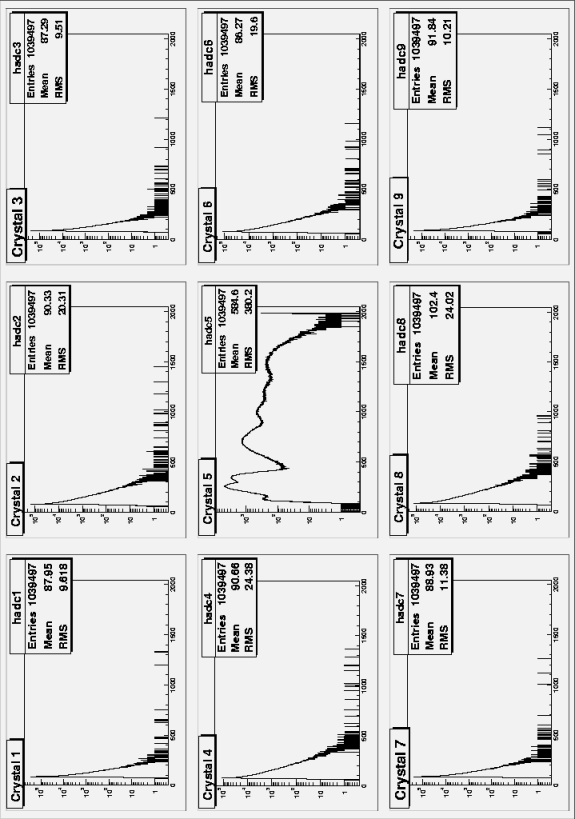}
\label{fig:9ch}}
\subfigure[Russian crystals]{
\includegraphics[width=0.6\linewidth,angle=270]{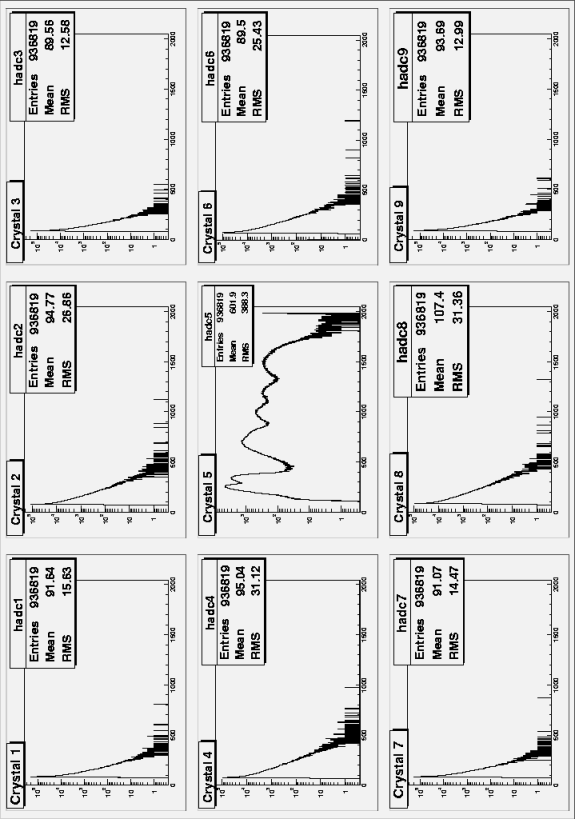}
\label{fig:9ru}}
\caption{The figures show the raw energy spectra corresponding to the
center crystals and the energy deposits in the surrounding crystals for
each array. It indicates that most of the energy is deposited into the
central element.}
\end{center}
\end{figure}

The spectra \ref{fig:9ch} and \ref{fig:9ru} illustrate the shower
deposition into the crystal arrays. Again, all eight selected photon
energies are superimposed.

\subsection{Relative Calibration}
The aim of the relative calibration is to put all peak positions for
all crystals at the same channel number, without knowing the absolute
energy. For this purpose, the absolute calibration must be performed
based on known reference values.

If all possible tagger channels would have been picked in the
experiment, a continuum would have been seen. Since only eight tagger
channels were selected, eight structures are seen in the energy
spectra. The locations of all eight peaks have to be determined in the
calibration process. Since the peaks overlap in the total energy
spectrum, they must first be separated. The separation is performed
using a method referred to as ``separation of peaks''. In this method,
the eight peaks in the energy spectrum are separated and instead of
having one spectrum with eight peaks, one will have eight different
spectra with eight different peaks. The separation of the peaks can be
done for each tagger energy separately by requiring a coincidence with
one particular tagger channel. This information has been stored as
well in the list mode data event-wise by using each tagger information
as a stop signal in the eight time measurements using the accepted
event as a start.  The peak positions in these eight individual energy
spectra are determined by fitting Gaussian functions.

The relative calibration method does not rely on the knowledge of how
much of the photon energy is deposited in the crystal or into its
neighbor, since all nine elements should show an identical response if
the photon beam is directed towards it center. The center crystal acts
as a reference since the total shower will later on be added to its
response. By comparing the eight different peak locations, with
pedestals subtracted, to the corresponding ones in the center crystal,
eight calibration factors can be deduced which should be identical
within statistical errors. The mean value is finally used as
normalization factor for the relative adjustment.

\subsection{Adding Energy Deposits Event-Wise}
The raw energy spectra taken from the 18 crystals show broad peaks
which partly overlap each other. By adding the energy contributions
from the eight surrounding crystals event-wise to the spectrum of the
center crystal, a more complete picture of the event is obtained. The
summed spectra have a much better energy resolution due to the added
energy and the processes in the crystals are described more properly
than before. 

The obvious difference between the raw spectrum and the summed
spectrum is, as expected, that the peaks are narrower since the photon
statistics has increased. Decreasing this uncertainty also decreases
the width of the peaks. Summing up the energy deposits causes the
counts in the tail of the peak to shift from lower energies to higher
energies. This further reduces the width of the peak.

The second observation is that the peaks have been slightly shifted to
higher energies. This effect comes directly from the adding of
energy. A peak corresponding to a high energy will of course be
shifted to the right in the energy spectrum. These higher energies
come much closer to the actual photon energy, but since the set-up is
not perfect and energy is lost in the material between the crystals,
the incoming photon energies are still higher. 

The third observation when comparing the raw and summed energy spectra
is that the peaks are separated to a greater extent after the
summation compared to before the summation. Even if the peaks are not
completely separated this is not really a problem in calculating the
energy resolution due to the available tagger information. In the
Russian energy spectrum of a single crystal one could distinguish six
peaks and after summing the spectra all eight peaks were visible. For
the Chinese crystals only seven peaks were seen after the summation
and this is an improvement compared to the raw spectrum where only
three peaks and a continuum were seen. The Russian crystals obviously
have a much better energy resolution since the peaks are narrower and
since all eight peaks are visible. That this is not the case for the
Chinese crystals is probably due mainly to three things; the light
yield of the crystals, their short length and their tapered shape. The
light yield of the Chinese crystals is only about 10 photoelectrons
per~MeV compared to about 20 for the Russian crystals. The shorter
length of the crystal will cause an increased energy leakage and
therefore a larger contribution to the resolution due to the
statistical fluctuations of the shower. In addition, the strongly
tapered shape will on one hand favor light collection at the larger
endface. However, the shape can impose a non-linear light collection
along the symmetry axis which contributes significantly to the
constant term in the energy resolution.  Of course, since all crystals
were individually wrapped in several layers of Teflon, aluminum foil
and shrinking tape, there is a substantial amount (about 0.4-0.5 mm
for each crystal) of dead material between the crystals which worsens
the energy resolution.

Looking closer on the peaks, one can see that they are not as
symmetric and Gaussian-shaped as would be expected. This effect is due
to the wrapping material and also to the fact that the showers may
spread out of the crystals and into the metal block or the surrounding
environment (nitrogen in this case). These two factors, or loss
mechanisms, influence the amount of detected energy; more losses means
less detected energy. The energy losses cause high energetic photons
to be interpreted as low energetic photons by the crystals. The result
is a shift of the peaks to lower energies and a tail on the left,
low-energetic side of the peaks, is formed. 

In neither of the two summed spectra, the peaks were completely
resolved and to clearly see the low energy tail, the peaks had to be
separated. A total separation is a very important feature when
treating the peaks. One must therefore first go through the separation
procedure using coincidence with the taggers before the peaks can be
analyzed and fitted with appropriate distributions.

\subsection{Calibration According to Incoming Energy}
After having performed the relative calibration all peak positions for
all crystals are identical. The peak positions in energy are not
known, though. To perform an absolute calibrations, simulations would
be needed since one does not know how much energy is lost in the
wrapping material or is leaking out of the detector. In the present
case one can simply calculate a conversion factor from channels to the
true incident energy, which however does not take into account the
true energy deposition. The calibration factor is obtained by the
ratio of the incoming energy in~MeV and the detected channel number
for this peak, see equation \ref{eq:calibration}. This has been done
for the experiments in Mainz. 

\begin{figure}[H]
\begin{center}
\subfigure[Chinese crystals]{
\includegraphics[width=0.5\linewidth,angle=270]{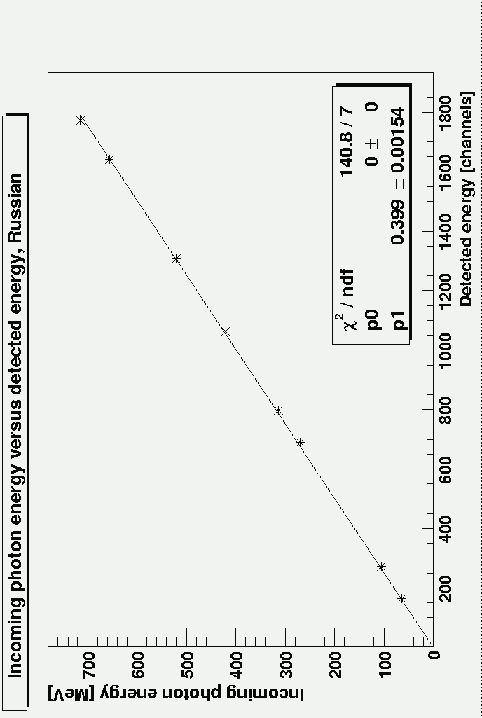}
\label{fig:slopezeroCH}}
\subfigure[Russian crystals]{
\includegraphics[width=0.5\linewidth,angle=270]{9spectraRU.png}
\label{fig:slopezeroRU}}
\caption{Figures of the incoming photon energy versus the detected
deposit in channel numbers for the eight energy peaks in the
Chinese and Russian summed energy spectra.}
\end{center}
\end{figure}

However, one should consider that the ratio between incident and
deposited energy does not have to be constant since the shower leakage
to the rear will increase with increasing photon energy. Nevertheless,
this does not seem to be a large effect since one sees in
\ref{fig:slopezeroCH} and \ref{fig:slopezeroRU} that it does not cause
a large deviation from linearity.

\begin{equation}
\label{eq:calibration}
f=\frac{E_s}{X_p}
\end{equation}

f is the so-called conversion factor that the counts in the spectrum
will be multiplied with in order for them to be expressed in~MeV
rather than in channels. $E_s$ is the incoming photon energy (in~MeV)
corresponding to a specific peak, and $X_p$ is the detected peak
position in channels. The detected peak position is strongly dependent
on what fit one chooses to use in order to obtain the peak positron
and on which interval one applies the fit on. However, this conversion
factor will not affect the obtained relative energy resolution
$\sigma/E$, since it is not dependent on what unit they are expressed
in as long as they are expressed in the same unit. 

Conversion factors are calculated for each peak in the summed
spectra. Finally the average value of these conversion factors are
used to change the scale from channels to~MeV.

\begin{table}[H]
\centering
\begin{tabular}{|l||c|c|}
\hline
Property &Chinese &Russian\\
\hline
\hline
Conversion factor & &\\
from channels to &0.4007 &0.3949\\
energy [MeV/ch] & &\\
\hline
\end{tabular}
\caption{The table shows the calculated conversion factors from
channels to energy. The values are calculated by dividing the energy
for a specific peak by the peak position in channels.}
\label{tab:conv}
\end{table}

This factor can be regarded as a preliminary conversion factor from
channels to~MeV. It neglects energy losses but can be still considered
as a good relative measure. To perform an absolute calibration, one
would need computer simulations to see how much energy is lost in the
system. The losses would typically result in yet another constant
factor by which the counts in the energy spectra were to be multiplied
with. 

\begin{figure}[H]
\begin{center}
\subfigure[Chinese crystals]{
\includegraphics[width=0.6\linewidth,angle=270]{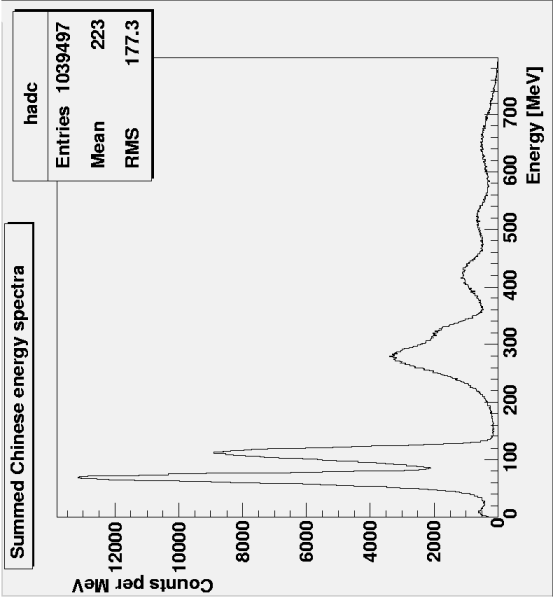}
\label{fig:summedegych}}
\subfigure[Russian crystals]{
\includegraphics[width=0.6\linewidth,angle=270]{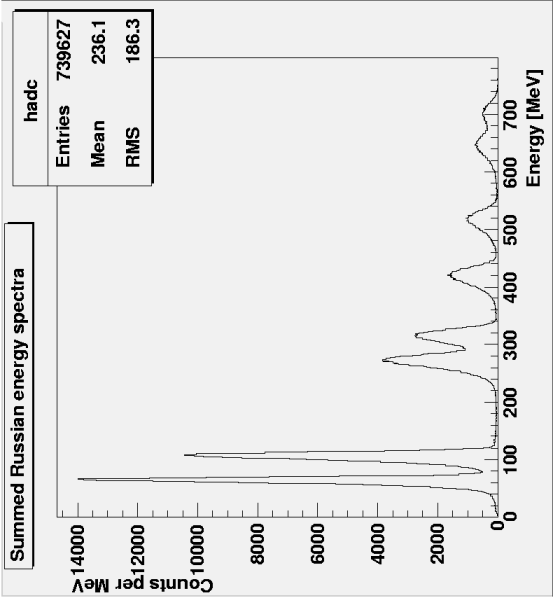}
\label{fig:summedegyru}}
\caption{Figures of the summed energy spectra for the Chinese and
Russian crystals. The energy scale is normalized to the incident energy.}
\end{center}
\end{figure}

Applying this factor to the peaks in the energy spectra,
\ref{fig:summedegych} and \ref{fig:summedegyru} are obtained.
\newpage

An important test that has to be done in connection to the relative
calibration is to investigate weather the energy response follows a
linear dependence with respect to the channel numbers. It is a general
test of the whole detector - linear response of the scintillation
process, the light collection as well as the photo multiplier response
- in addition also a check of the calibration procedure, the errors in
the method and the deduced conversion factors.  By plotting the
incident photon energy versus the detected energy in arbitrary
channels, one should obtain a straight line, and its slope ought to be
identical to the conversion factors calculated above.

The deviations from linearity in the graphs \ref{fig:slopezeroCH} and
\ref{fig:slopezeroRU} are less than 5~MeV for the peak position at 700
~MeV (0.71\%) for the Russian crystals and less than 10~MeV for the
peak position at 272~MeV (3.68\%) for the Chinese ones. These
deviations are as well seen in table \ref{tab:dev}.

\begin{table}[htb]
\centering
\begin{tabular}{|l||c|c|}
\hline
Crystal &Deviation from &Deviation from \\
&linearity [MeV] &linearity [\%]\\
\hline
\hline
Chinese &10 &3.68\\
\hline
Russian &5 &0.71\\
\hline
\end{tabular}
\caption{The table shows the maximum deviation from linearity in
figures \ref{fig:slopezeroCH} and \ref{fig:slopezeroRU}. The deviation
was measured at the energies 272~MeV and 700~MeV for the Chinese and
Russian crystals respectively.}
\label{tab:dev}
\end{table}

As shown in \ref{tab:compare}, the slopes of the graphs do not
coincide with the calculated conversion factors. This is due to the
fact that when calculating the average conversion factors, all
calculated factors are given equal weight. This is not the case when
performing the straight line fit. Here, the data point representing
the highest energy restricts the possible values of the conversion
factor more than the other data points (and especially the data point
representing the lowest energy). Hence in this case, the data points
have been given the same weight factors and data points far from the
origin influence the conversion factor stronger.

\begin{table}[htb]
\centering
\begin{tabular}{|l||c|c|}
\hline
 &Chinese crystals &Russian crystals \\
\hline
\hline
Calculated factor &0.4007 &0.3949\\
\hline
Slope of graph &0.4134 &0.3990\\
\hline
Difference &0.0127 &0.0041\\
\hline
Difference in \% of &3.2 &1.0\\
the calculated factor & &\\
\hline
\end{tabular}
\caption{The table shows the conversion factors obtained when using
the calculated factor and the slopes from the linearity graphs. The
difference between the values is greater in the case of the Chinese
crystals.}
\label{tab:compare}
\end{table}

\section{Resolution Dependence on Photon Energy}
The energy resolution depends on the energy of the incoming
photons. The dependence is depicted by plotting the sigma of the peaks
divided by the incident energy, versus the incoming photon energy
given in~GeV. The energy dependence can be parametrized by equation
\ref{eq:res}.

\begin{equation}
\label{eq:res}
\frac{\sigma}{E}=\frac{p_1}{\sqrt{E(GeV)}}+p_0
\end{equation}

\subsection{Different Methods of Fitting the Peaks}
\label{diff}
When calculating the energy resolution, different methods can be used
to obtain the $\sigma/E$. These methods differ in how to fit the
energy peaks, or at least in how to obtain the peak position, the
width of the peak and as well other characteristic properties. 

The obvious choice in obtaining a measure of the width of the peaks,
the $\sigma$-value, is to take the FWHM (Full width Half Maximum) of
them. The FWHM is used whenever one is dealing with non-Gaussian
shaped peaks. Although the peaks in the Russian and Chinese energy
spectra look Gaussian, the low energy tail destroys the symmetry. The
disadvantage with this method is that the FWHM value will be
unnecessarily large and that it will not describe the width
properly. This is easily understood if one thinks of the low energy
tail as created due to energy losses. These losses will increase the
peak width and worsen the energy resolution so that the energy
resolution looks worse than it actually is. The FWHM value is obtained
by registering the height of the separated peaks in the spectra and
writing down the width of the peaks at half of this maximum.

Another method has therefore been used to determine the energy
resolution of the crystals. This method is based on the assumption
that the separated asymmetric energy peak contains two sets of samples
that can be described by two different Gaussian functions. One of
these functions represents events where no energy escapes from the
$3\times3$ array of crystals; this curve is called the signal
peak. The second curve makes up for the difference between the signal
peak and the real asymmetric peak in the energy spectrum. The second
peak physically represents the events where part of the energy is
lost; that is the energy that stays in the material between the
crystals or leaks out of the system in other ways. By adding these two
Gaussians an asymmetric function, looking much like the original
energy peak, is obtained.

\begin{figure}[H]
\centering
\includegraphics[width=0.6\linewidth,angle=270]{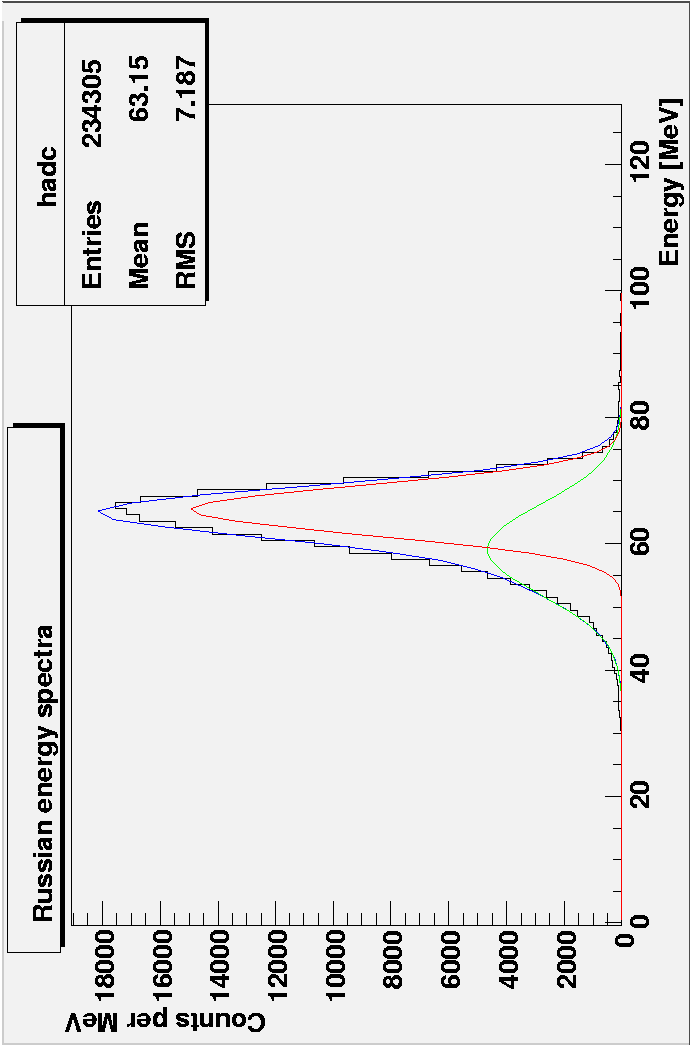}
\caption{The figure shows the two different curves used to describe
the counts belonging to the signal and the lost energy
respectively. These curves are red and green respectively. The sum of
the two Gaussians is as well displayed (blue).}
\label{fig:rusepfit_lowegy}
\end{figure}

The method with the two Gaussian functions is applied to both the raw
and summed energy spectra. The difference between the sigma values in
the two cases affects the energy resolution. By adding the energy
contributions, the peaks get narrower, the sigma decreases and the
energy resolution improves.

To compare the two widths of the peaks (the FWHM and the $\sigma$) one
can divide the FWHM with 2.35. A corresponding $\sigma$-value will
then be obtained. The comparison shows that the sigma value calculated
from the FWHM value in most of the cases is much larger than the sigma
value obtained from the signal peak. This is expected, since the
asymmetric peak obviously is much wider than the signal peak.

\subsection{Right Side Fit}
\begin{figure}[H]
\begin{center}
\includegraphics[width=0.6\linewidth,angle=270]{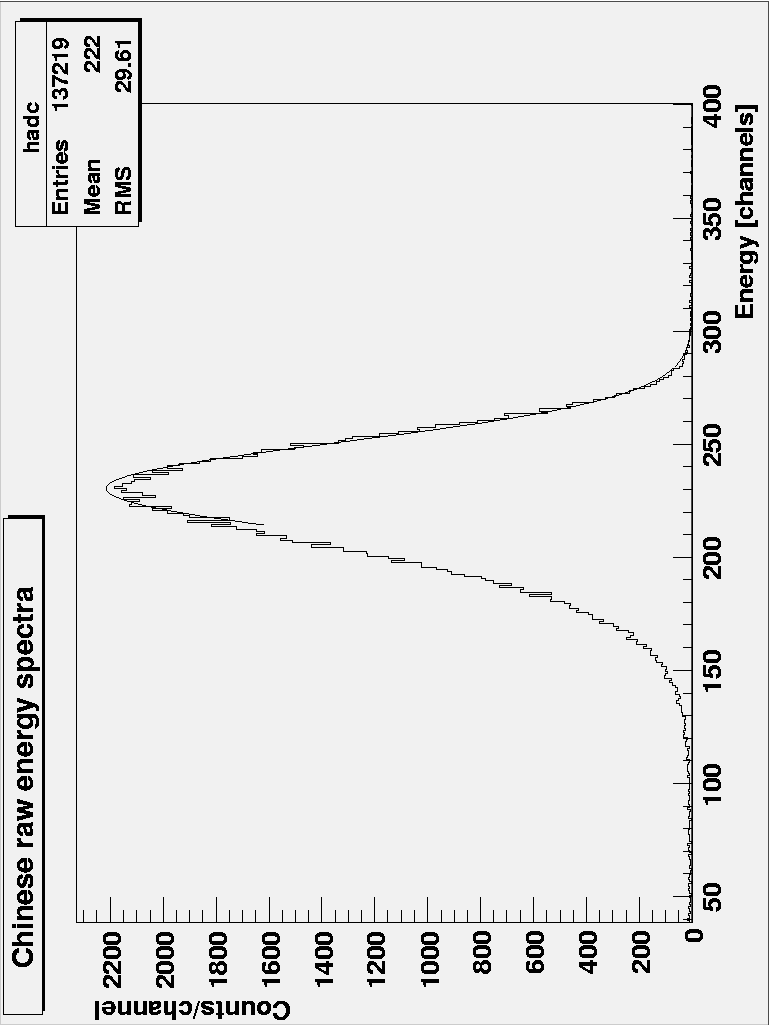}
\end{center}
\caption{The figure shows how the Gaussian fits were performed. Due to
the asymmetry and the tail to the left of the peak, only the right
side of the peak was fitted.}
\label{fig:showfit}
\end{figure}

Because of the unreliability (or unlikeliness) in the energy
resolutions obtained from the FWHM and the signal peak, it is
important to fit the energy peaks with a more appropriate
function. One could simply pretend that they have the shape of a
Gaussian distribution and ignore the low energy tail. According to
this method, all Gaussian fits are carried out over an interval that
ranges from the upper left part of the peak to the end of the peak on
the right side as shown in \ref{fig:showfit}.

This fitting procedure may initially seem strange and not very
reliable. It is however based on the assumption that some of the
energy is lost in the material between the crystals. This shifts some
of the counts in the energy spectra to lower energies and creates a
tail on the left side of the peak. If no energy would be lost, one
could expect this tail to disappear but the right side of the
fit would still look the same. 

Repeating the fitting procedure shown in figure \ref{fig:showfit} for each
and every one of the separated summed energy peaks yields yet another
measure of the peak width. Plotting the $\sigma/E$ against the energy
in~GeV gives an energy resolution somewhere in between the two
extremes discussed in section \ref{diff}.

\begin{figure}[H]
\begin{center}
\subfigure[Chinese crystals]{\includegraphics[width=0.6\linewidth,angle=270]{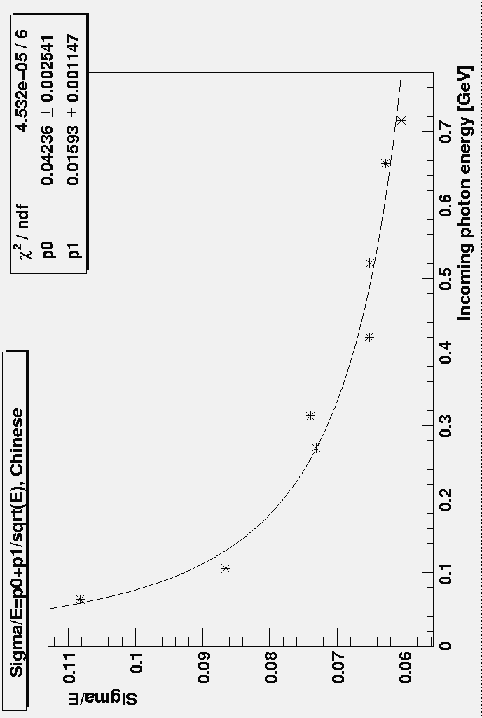}
\label{fig:sigdivECHgig}}
\subfigure[Russian crystals]{\includegraphics[width=0.6\linewidth,angle=270]{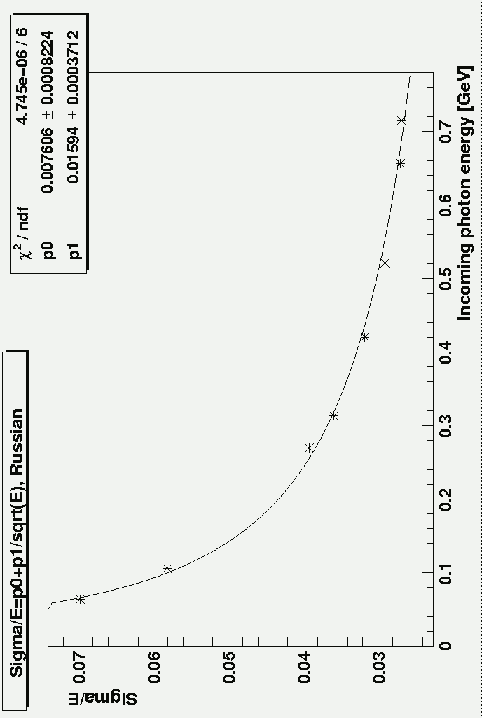}
\label{fig:sigdivERUgig}}
\caption{The figures shows the plot of the sigma divided by the energy
as describes above for the Chinese and Russian crystals. The energy
resolution is described by a polynomial of the form
$\frac{p_1}{\sqrt{E}}+p_0$.}
\end{center}
\end{figure}

Using this method, the values of $p_0$ and $p_1$ for the Russian
crystals come even closer to the desired values given in the proposal
for the ECAL, $\frac{1.54\%}{\sqrt{E}}+0.3\%$, see table
\ref{tab:egyres} for comparison. The values of the coefficients were
determined by the fit to be 4.24\% and 1.59\% for the Chinese crystals
and 0.76\% and 1.59\% for the Russian crystals.

\begin{table}[htb]
\centering
\begin{tabular}{|l||c|c|}
\hline
Crystal &$p_0$ in \% &$p_1$ in \%\\
\hline
\hline
Chinese &4.24 $\pm$ 0.25 &1.59 $\pm$ 0.11\\
\hline
Russian &0.76 $\pm$ 0.08 &1.59 $\pm$ 0.04\\
\hline
\end{tabular}
\caption{The table shows the resulting coefficients in the function
describing the energy resolution for the crystals. The energy
resolution for the Chinese crystals is given by equation \ref{eq:res}.}
\label{tab:egyres}
\end{table}

The maximum deviations from the functions in \ref{fig:sigdivECHgig}
and \ref{fig:sigdivERUgig} describing the energy dependence of the
relative energy resolution are less than 0.3\% for the Chinese
crystals and an energy of 313.3~MeV and less than 0.1\% for the
Russian crystals and an energy of 269.8~MeV.

\section{Resulting Energy Resolution}
\subsection{Comparing the Three Methods}
\begin{figure}[H]
\begin{center}
\subfigure[Chinese crystals]{
\includegraphics[width=0.6\linewidth,angle=270]{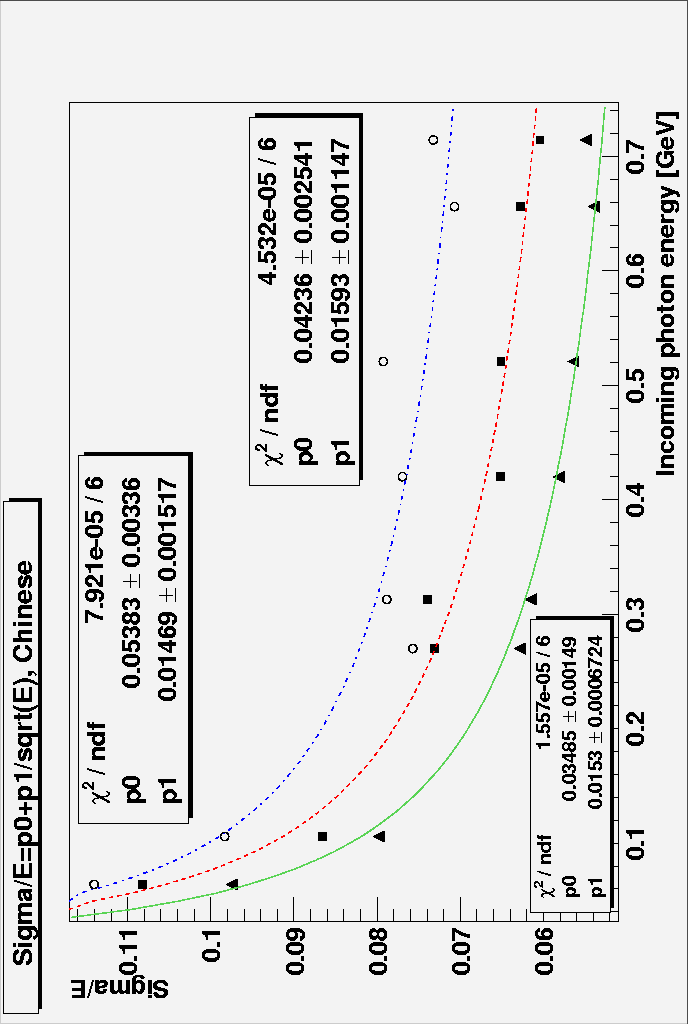}
\label{fig:sigdivECHcomparison}}
\subfigure[Russian crystals]{
\includegraphics[width=0.6\linewidth,angle=270]{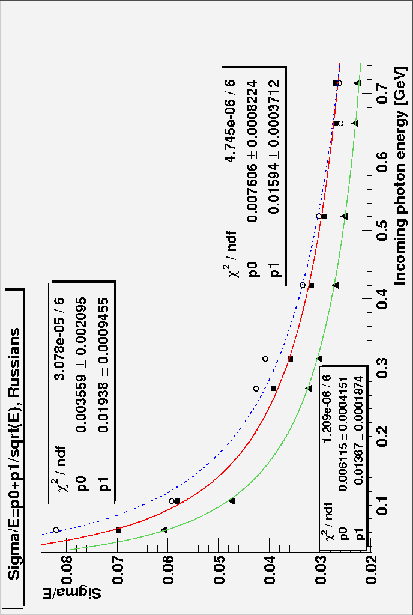}
\label{fig:sigdivERUcomparison}}
\caption{The figure shows the energy resolutions from the three
 methods discussed above. The curve with open circles shows the energy
 resolution one gets when using the FWHM method. The curve with closed
 boxes shows the energy resolution obtained when using the right side
 fit and the curve with closed triangles shows the energy resolution
 achieved when using the signal peak from the summed energy spectra.}
\end{center}
\end{figure}

In order to see which of the three methods gives the lowest energy
resolution (lowest values of $p_0$ and $p_1$), one must compare the
results from the calculations. This comparison is seen in
\ref{fig:sigdivECHcomparison} and \ref{fig:sigdivERUcomparison}. Here,
one can see that the energy resolution obtained when using the sigma
from the signal peak gives the lowest resolution. This case may be an
indication of which resolution may be achieved if performing very good
measurements with a sufficient amount of crystals absorbing the
energies from the incoming photons and with a minimum of dead material between
the crystals.
\newpage

The FWHM and the signal peak can be seen as two extreme scenarios that
are not very likely to occur, although they may happen. The right side
fit seems to give the best description of the energy resolution since
the result is somewhere between the other two. 

\subsection{Final Energy Resolution}
In order to give an estimation of the systematic uncertainty in the
results obtained with the right side fit, half of the deviation in
energy resolution between this method and the other two is taken as
$\sigma$. For the Chinese crystals, the systematic shift of the three
curves in \ref{fig:sigdivECHcomparison} indicate that the $\sigma$
should be almost constant. For the Russian crystals, the $\sigma$ is
larger for lower energies than for higher since the FWHM and the right
side fit gave almost the same asymptotic energy resolution.

\begin{figure}[H]
\begin{center}
\includegraphics[width=0.6\linewidth,angle=270]{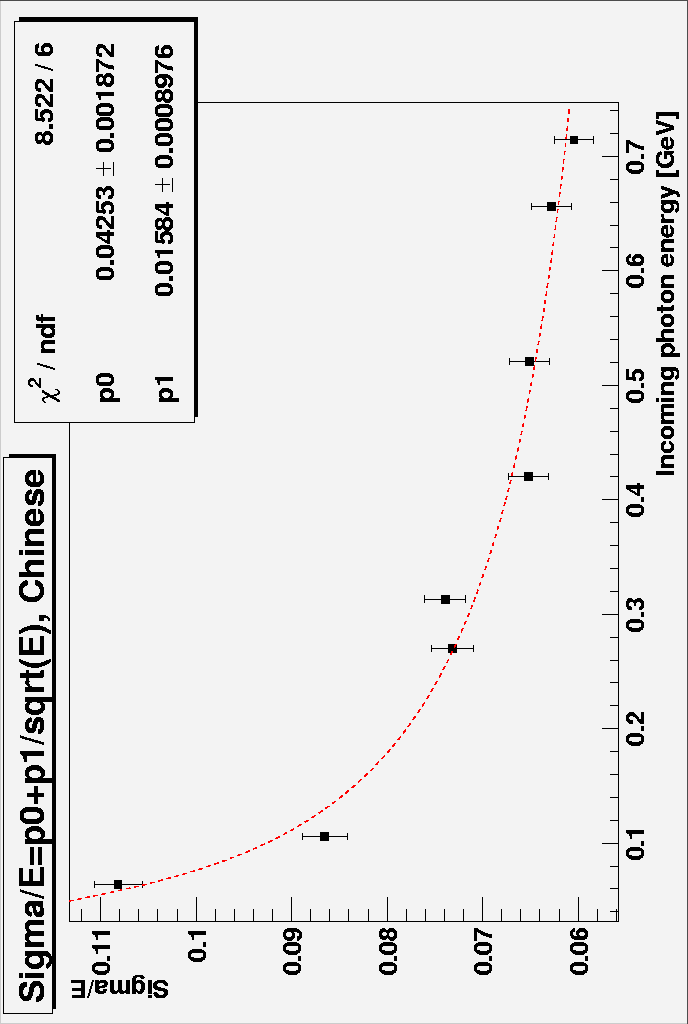}
\label{fig:cherrorbar}
\caption{The figures show the energy resolutions obtained for the
Chinese crystals with the right side fit and the systematic
uncertainty in the data points. The uncertainty is based on the three
different energy resolutions obtained with the three different fitting
methods.}
\end{center}
\end{figure}
\newpage

\begin{figure}[H]
\begin{center}
\includegraphics[width=0.6\linewidth,angle=270]{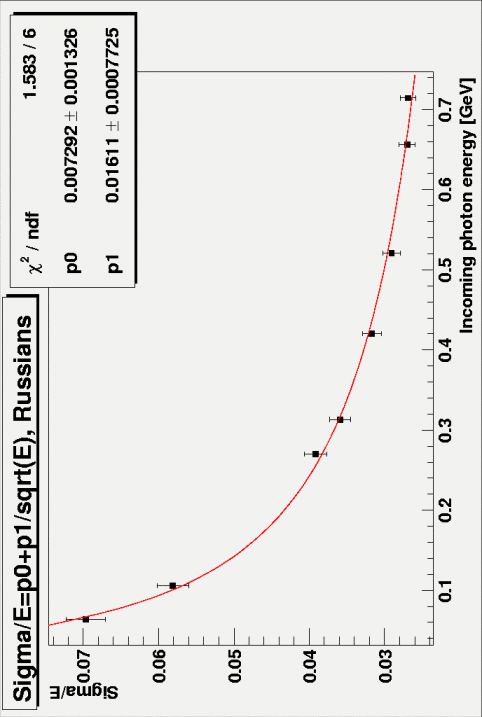}
\label{fig:ruerrorbar}
\caption{The figures show the energy resolutions obtained for the
Russian crystals with the right side fit and the systematic
uncertainty in the data points. The uncertainty is based on the three
different energy resolutions obtained with the three different fitting
methods.}
\end{center}
\end{figure}

\begin{table}[htb]
\centering
\begin{tabular}{|l||c|c|}
\hline
Final energy resolution &$p_0$ &$p_1$ \\
\hline
\hline
Chinese crystals &4.25 $\pm$ 0.19 &1.58 $\pm$ 0.09\\
\hline
Russian crystals &0.73 $\pm$ 0.13 &1.61 $\pm$ 0.08\\
\hline
\hline
CDR Report &0.3 &1.54\\
\hline
\end{tabular}
\caption{The table shows the final energy resolution obtained when
using the right side fit and error bars describing the systematic
uncertainties in the results. The energy resolution is described with a
polynomial $p_1/\sqrt{E}+p_0$. The uncertainty in $p_0$ and $p_1$ above
are uncertainties in the fitting, not a measure of the size of the
error bars.}
\label{tab:resolution}
\end{table}

The error bars for the Chinese crystals in \ref{fig:cherrorbar} range
from $\pm$0.21\% for low energies to $\pm$0.25\% for high
energies. The error bars for the Russian crystals in
\ref{fig:ruerrorbar} range from $\pm$0.10\% for low energies to
$\pm$0.26\% for high energies.\\

\begin{table}[htb]
\centering
\begin{tabular}{|l||c|c|}
\hline
Final energy resolution &Minimum error bar &Maximum error bar \\
\hline
\hline
Chinese crystals &$\pm$0.25~\% &$\pm$0.21~\%\\
\hline
Russian crystals crystals &$\pm$0.26~\% &$\pm$0.10~\%\\
\hline
\end{tabular}
\caption{The table shows the final energy resolution obtained when
using the right side fit and error bars describing the systematic
uncertainty in the results. The energy resolution is described with a
polynomial $p_1\sqrt{E}+p_0$. The uncertainty in $p_0$ and $p_1$ above
are uncertainties in the fitting, not a measure of the size of the
error bars.}
\label{tab:errresolution}
\end{table}

\section{Dependence of the Energy Resolution on the Array Size}
\begin{figure}[H]
\begin{center}
\includegraphics[width=0.5\linewidth,angle=270]{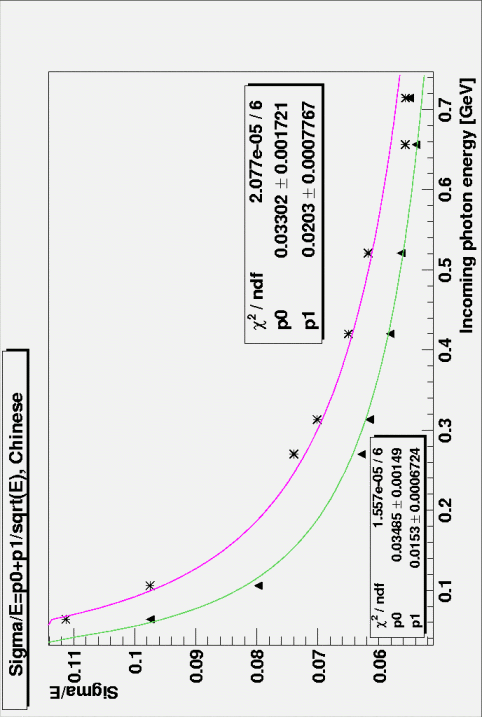}
\label{fig:sigdivECHsumunsum}
\caption{The figure shows the graph describing the energy resolution
of the raw and summed Chinese energy spectra. The curve marked with
stars describes the energy resolution for the raw spectra. The
constant p1 is lower for the summed spectra, this indicates a lower
energy resolution. The fitting of the energy peaks has been performed
with the two Gaussian method and the $\sigma$ comes from the signal
peak.}
\end{center}
\end{figure}

\begin{figure}[H]
\begin{center}
\includegraphics[width=0.5\linewidth,angle=270]{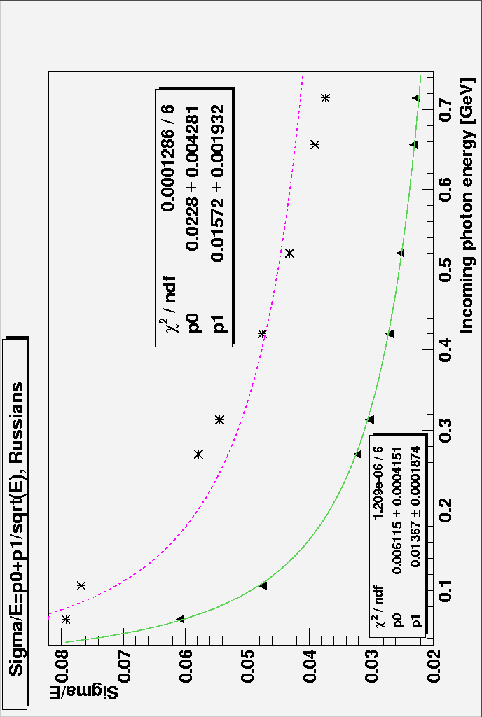}
\label{fig:sigdivERUsumunsum}
\caption{The figure shows the two graphs describing the energy
resolution of the raw and summed Russian energy spectra. The curve
marked with stars describes the energy resolution for the raw
spectra. The constant p1 is lower for the summed spectra, this
indicates a lower energy resolution. The fitting of the energy peaks
has been performed with the two Gaussian method and the
$\sigma$ comes from the signal peak.}
\end{center}
\end{figure}

One as well may want to know how much better it is to use a large
array of crystals compared to using only a small one. This effect can
(in some way) be seen by comparing the energy resolutions obtained
from the raw and summed energy spectra. These energy resolutions
differ a lot. The two curves describing the resolution can be seen in
\ref{fig:sigdivECHsumunsum} and \ref{fig:sigdivERUsumunsum}. In both
figures, the energy resolution is obtained from the signal peak.

\newpage

Not summing over the energy contributions in the surrounding crystals
obviously gives a worse energy resolution since $p_0$ and $p_1$ have
clearly lower values. This is shown in the graphs.

\section{Conclusions from the Experiment at MAMI}
Using three different methods of estimating the energy resolution
of the lead tungstate crystals gave three quite different results. The
largest energy resolution was obtained using the FWHM value of the
peaks. This was expected since only parts of the peaks were used to
estimate the energy resolution in the two other cases. The lowest
energy resolution was obtained when using the sigma of the signal
peak. The assumption to use only the right side of the energy peak to
fit of a Gaussian distribution led to an energy resolution between the
two other extreme values. 

The energy resolution obtained from the FWHM can be seen as a worst
case scenario for the energy resolution since the tail of the peak
influences and increases the FWHM value and hence worsens the
resolution. The signal peak on the other hand, can be seen as the
ideal case where no energy is lost and the energy resolution indicates
what may be achieved with a perfect set-up and instrumentation of a
high quality. 

The energy resolutions obtained for the Russian crystals, taking into
account the systematic uncertainties,
$\frac{1.61\%}{\sqrt{E(GeV)}}+0.73\%$ came close to what was proposed
in the Conceptual Design Report report about the crystals to be used
at GSI ($\frac{1.54\%}{\sqrt{E(GeV)}}+0.3\%$). The Chinese crystals,
seem to have a worse energy resolution
($\frac{1.58\%}{\sqrt{E(GeV)}}+4.25\%$) dominated by the constant
term. 

Due to the difference in geometry, a direct comparison is not
possible. The statistical factors are very similar both for the
Chinese and Russian crystals, which indicates similar luminescence
properties. However, the coefficient before the $1/\sqrt{E}$-term is
approximately the same as for the Russian crystals, especially if one
considers the uncertainty in this value. The term $p_1$ represents
light output of the crystals and photon statistics and it should be
smaller for the Russian crystals due to the higher light yield. The
term $p_0$ describes the performed calibration, the set-up properties and
time dependent parameters such as temperature or count rate changes or
the various effects due to light collection or shower leakage. Due to
the tapered shape of the Chinese crystals, a larger part of the shower
escapes and thus the leakage is larger and the value of this term
increases. 

Although the result came close to the desired energy resolution stated
in the CDR, the energy resolution of the crystals was not as good as
expected. Previous measurements of $PbWO_4$ crystals, conducted in
1997 in Giessen, Germany \cite{diploma}, resulted in an energy
resolution of $\frac{1.69\%}{\sqrt{E(GeV)}}+0.63\%$. This value was
obtained for a $5\times5$ array of crystals, but without any cooling
of the crystals. Improvements have been done since then, both in the
light yield of the crystals and in cooling techniques. The light yield
has improved by a factor of two and the cooling has improved the light
yield with a factor of three. That should result in an
energy resolution that is a factor $\sqrt{6}$ lower than for the
previous measurement. This seems not to be the case and it is
therefore very important to repeat the measurements. The crystals
should then not be individually wrapped and one must ensure an
excellent optical contact between the crystals and the read-out
devices. Also a proper PMT set-up should be used in order to collect
as many photons as possible. In addition one must verify, or
optimize, the linear response of each detector by measuring the
response to low energy gamma sources as a function of position along
the axis of the crystal.

\cleardoublepage
\chapter{Summary and Outlook}
\section{Summary}
The objective of this thesis was to determine the energy resolution
of two sets of recently developed crystals. The crystals will be used
to detect gammas (and electrons) in an Electromagnetic Calorimeter at
the future upgrade of the GSI facility.

The gammas used in the experiment ranged in energy between 64~MeV and
715~MeV. The crystals were irradiated by eight photon energies in
this interval and the signals from the crystals were read out using
PMTs. 

The energy spectra obtained in the experiment showed eight peaks, one
corresponding to each photon energy used. The peaks were all
asymmetric with a tail to lower energies. Due to the asymmetry, the
sigma of the peaks was obtained using three different
methods. Firstly, the FWHM of the peak was used. Next, two different
Gaussian distributions were fitted to the peak in order to describe
the asymmetry. One of them represented events where no energy escapes
the detector while the other one represented the energy lost in the
wrapping material and as well the energy leaking out of the
detector. At last, a Gaussian function was fitted to mainly the right
side of the peak.

The sigma from the FWHM turned out to give the largest energy
resolution while the sigma from the signal function gave the
lowest. The right side fit resulted in an energy resolution somewhere
in between the other two values. 

The use of two sets of crystals with different shape and light yield
resulted in two different sets of results of the energy
resolution. Using crystals with a length of approximately 17 radiation
lengths and a higher light yield gave a lower energy resolution. The
energy resolution from the right side fit for the Russian crystals was
estimated to be $1.61\%/\sqrt{E}+0.73\%$, this value comes very close
to the value written in the Conceptual Design Report,
$1.54\%/\sqrt{E}+0.3\%$. The Chinese crystals were found to have an
energy resolution of $1.58\%/\sqrt{E}+4.25\%$.  

Using the FWHM of the peaks gives a very bad resolution. Using the
signal peak gives a very low and desirable resolution but it is
somehow unclear how one could obtain such a narrow and symmetric
energy peak. The right side fit seems to be most appropriate since the
energy value obtained using that method is between the two
extremes. The result in this case is significantly improved compared
to measurement using crystals of minor quality. In spite of a smaller
volume the achieved resolutions are very close to values obtained for
a $5\times5$ matrix with identical read-out but operated at higher
temperatures. Nevertheless, the improvement of the light yield of a
factor two compared to former crystals, which is valid for the Russian
as well as the Chinese samples of 15 cm length, as well as the
operation at -24~degrees~Celsius should have led to an even better
results. One should have expected an increase of the light yield by a
factor $\sqrt{6}$, which should much further reduce the statistical
factor $p_1$ in the parametrization of the relative energy resolution.

\section{Outlook}
The final energy resolution of the crystals was obtained using a
method ignoring the low energy tail of the energy peaks. To really
understand and explain the existence of this tail, simulations would be
needed. It is clear that the dead material is partly responsible for
it, but other explanations are needed as well. The simulations may also
provide an opportunity to verify the use of more than one function to
describe the shape of the energy peaks. Possibly, they could also
support the decision to use the two Gaussian method and as well
justify the assumption that the signal peak may be the ideal energy
resolution to strive for. Another important reason for conducting
simulations would be to investigate exactly what size of the crystal
array is needed to collect all the showers in the detector.

Simulations in these areas were begun but due to lack of time, no
proper results were obtained. The only conclusion one could draw from
those simulations was that an array larger than the $3\times3$ array
used in the experiments was needed. This was especially important for
photon energies of high energy, when deposits in the surrounding
crystals were larger. The thickness of the wrapping layer was as well
varied in the simulations to investigate whether this influenced the
amount of energy deposits. No conclusions could be drawn in this case
since the Z-value of the wrapping material was so much lower compared
to that of the crystals. This means that the wrapping material is not
capable of collecting very many photons and that they instead escape
the detector. Hence this factor did not noticeably affect the
depositions.

Since the energy resolution obtained in the experiment was good, but
not as good as expected due to the improvements made since the last
time the experiments were performed, new experiments are required. In
the new experiments it is important to minimize the amount of dead
material in the detector and if possible, exclude the individual
wrapping of the crystals. It is also important to assure a good
optical contact between the crystals and the read-out devices and to
use a proper PMT set-up that is more stable than the one used in
Mainz. This will most likely increase the amount of detected photons
and improve the energy resolution of the crystals.

As the crystal manufacturing develops, crystals with a higher light
yield than the one used in this experiment are expected in the near
future. This will drastically impact the energy resolution. 

It has, at a late stage of this thesis, come to my knowledge that the
asymmetry of the peaks could be better described with the so-called
Novosibirsk function, rather than with two Gaussian
distributions. This function looks like a Gaussian distribution but
has a low energy tail. Applying it to the fit of the energy peaks may
very well yield a better description of the energy resolution and it
would therefore be best to repeat the fitting with this function.

\cleardoublepage
\chapter*{Acknowledgment}
I would like to thank Rainer Novotny at the Justus Liebig University
in Giessen for giving me the opportunity to participate in the
experiments in Giessen and Mainz and for answering my never-ending
questions with such patience. He has also been very helpful with
giving me comments and feedback on my report. 

I would also like to thank Inti Lehmann at Uppsala University for his
good ideas all through this thesis and I would as well like to
thank him for helping me structuring my report. 

Also thanks to Henrik Pettersson, Uppsala University, for helping me
overcome computer problems and increasing my knowledge of computer
simulations. 

Finally, thank you Ulrich for providing me the opportunities to conduct
my master thesis in such an interesting way and for letting me present
my results at the collaboration meeting at GSI. Your encouragement has
been a true inspiration.

\end{document}